\documentclass[aps,prb,twocolumn,floatfix]{revtex4-1}
\usepackage{graphics}
\usepackage{subfigure}
\usepackage{epsfig}
\usepackage{epsf,epic}
\usepackage{color}
\usepackage{marvosym }
\usepackage{subfigure}
\usepackage{amsmath}
\usepackage{amssymb}
\usepackage{amsfonts}
\usepackage{wrapfig}
\usepackage{multirow}
\usepackage{bm}
\usepackage{dcolumn}
\newcommand{\etal}{\textit{et al.\ }}
\newcommand{\ie}{\textit{i.e.\ }}

\begin{document}
\title{Computational study of Electron Paramagnetic Resonance  parameters for Mg and Zn impurities in $\beta$-Ga$_2$O$_3$}
\author{Dmitry Skachkov and Walter R. L. Lambrecht}
\affiliation{Department of Physics, Case Western Reserve University, 10900 Euclid Avenue, Cleveland, OH-44106-7079, U.S.A.}

\begin{abstract}
  A  computational study of the electron paramagnetic  resonance (EPR) $g$-tensors  and hyperfine tensors   in Mg and Zn doped $\beta$-Ga$_2$O$_3$ is presented.
  While Mg has been found previously to prefer the octahedral site, we find here that Zn prefers the tetrahedral substitutional site. The EPR signatures are
  found to be distinct for the two sites. Good agreement with experiment
  is found for the $g$-tensor and hyperfine interaction for Mg$_\mathrm{Ga2}$
  and predictions are made for the Zn case. 
\end{abstract}
\maketitle

\section{Introduction}
Monoclinic $\beta$-Ga$_2$O$_3$ with a band gap of
about 4.7$\pm$0.1 eV,\cite{Matsumoto74,Peelaers15a,Furthmuller16, Mengle16,Ratnaparkhe17} has recently
attracted attention as an ultra-wide-band-gap semiconductor
for transistor and transparent conducting applications.\cite{Sasaki13,Green16}
As for any new semiconductor, a thorough understanding of the defects and
dopants and ultimately their control is crucial to the development of this
material. Several previous theoretical works have addressed the point
defects.\cite{Varley11,Varley12,Harwig78,Zacherle13,Peelaers16,Deak17,Kyrtsos17,Kyrtsos18}
On the experimental side, Electron Paramagnetic Resonance (EPR)
provides one of the most powerfull methods to identify the chemical nature
of defect centers. Several papers recently reported EPR centers in
$\beta$-Ga$_2$O$_3$.\cite{Kananen17,Kananensth,KananenMg,Jurgen18,Ho18,Skachkov19}

In this paper we address the recently reported Mg$_\mathrm{Ga}$
acceptor type dopant.\cite{KananenMg}
We present first-principles calculations of
the $g$-tensor and hyperfine parameters characteristic of this defect,
which support the  previous assignment of this defect with the Mg
on the octahedral Ga site.  In fact, we present calculations for both
sites and show that they would be distinctly different.  Previous
computational work\cite{Ho18} indeed finds Mg to have lower energy on the
octahedral site. Encouraged by this successful agreement with experiment,
we then consider another candidate acceptor, Zn$_\mathrm{Ga}$ and
predict its EPR signatures.  Our reason for choosing Zn is that Zn
may prefer the tetrahedrally coordinated site. In fact, Mg in MgO occurs
in a rocksalt structure with octahedral environment but Zn in ZnO has a
tetrahedral bonding.  Thus we anticipated that Zn might prefer the
tetrahedral site in $\beta$-Ga$_2$O$_3$.  We will show that this
hypothesis is confirmed and predict the corresponding $g$-tensor
and hyperfine splittings. 

\section{Computational Method}
The $g$-tensor is calculated using the Gauge Including Projector
Augmented Wave (GIPAW) method.\cite{Pickard01,Pickard02,Gerstmann10,Ceresoli10} 
This is  a Density Functional Perturbation Theory (DFPT) method
to calculate the linear magnetic response of a periodic system
onto  an external magnetic
field. It is implemented in the code QE-GIPAW,\cite{gipaw} which is integrated within the Quantum Espresso package.\cite{QE-2009}

The hyperfine tensor calculation
is also incorporated in the GIPAW code although it does not strictly
require the GIPAW methodology. The hyperfine tensor has two parts: 
the isotropic
Fermi contact term which depends on the wave function or spin density
of the defect at the nuclear sites of atoms with a net nuclear spin
and a dipole interaction term which is non-isotropic.  The hyperfine
interaction is sensitive to the degree of localization of the
defect wave fuctions. The latter tends to be underestimated by the
local density approximation (LDA)  or even generalized gradient approximation
(GGA) to the exchange-correlation functional, particularly
for acceptors. This is because the latter does not fully cancel the Coulomb
self-interaction. This can in part be remedied by using an orbital
dependent functional such as a hybrid functional or in a less expensive manner using DFT+U, in which on-site orbital specific Coulomb interactions are added.
These have the effect of shifting empty states (hole states) up in energy
and deeper in the gap, thereby making them more localized. Typically,
this also involves a feedback in the relaxation of the structure which then
tends to become localized on a single atom instead of spreading of the
several nearest neighbors of a defect site.

In the case of $\beta$-Ga$_2$O$_3$
the acceptor states tend to be localized on O-$p$ like dangling bonds
because these comprise the top of the valence band. In previous works\cite{Jurgen18,Skachkov19}
on the EPR parameters of Ga-vacancy related states, we found it 
was necessary to apply a rather large $U$ on O-$p$ states, in order to
obtain adequate localization on their neighboring O atoms and reduce
the $s$-like spin density on the second neighbor which determines
the superhyperfine (SHF) interaction. 
Within pure semi-local functionals, the SHF interaction on the
Ga neighbors to the O on which the spin density becomes localized
was overestimated by almost a factor 2 even if the structure was relaxed
with DFT$+U$ or hybrid functional. 
On the other hand, Ho \etal\cite{Ho18} showed that using a
hybrid functional, the Mg$_\mathrm{Ga}$ acceptor ---which has its spin density
localized on a nearby O and is thus in many ways similar to the vacancies---
good agreement was obtained between theory and experiment for the
strength of the hyperfine coupling.  Here we will show that this
is the case also in   pure GGA provided the structure is relaxed using GGA+U.
This is  because the defect spin density is then already
very well localized on a single O and not as sensitive to the $U$ value. 

At present, the GIPAW code does not yet allow us to calculate the $g$-tensor
using electronic structures at the DFT+U or hybrid functional level. Thus
our $g$-tensor calculations are performed at the GGA level using
the Perdew-Burke-Ernzerhof\cite{PBE} functional but using structures relaxed
in DFT+U.

\begin{table*}
  \caption{Calculated EPR parameters for Mg$_\mathrm{Ga}$ and Zn$_\mathrm{Ga}$ defects. In our results the $g$ tensor and $A$-tensors
    are given in terms of three principal values followed by the $\theta$ (polar)  and $\phi$ (azumuthal) angles in degrees
    measured from ${\bf b}$ and ${\bf a}$ respectively.
    The results of other groups pertain to Mg$_\mathrm{Ga2}$ and
    are given with respect to ${\bf a}_*$, ${\bf b}$ and ${\bf c}$  axes,
    which correspond to $\theta=90^\circ,\phi=14^\circ$, $\theta=0^\circ$, $\phi$ arbitrary and $\theta=90^\circ,\phi=-76^\circ$ respectively.
       \label{tabepr}
   }
  \begin{ruledtabular}
    \begin{tabular}{l|l|ccc|ccc|ccc|ccc}
      Model             &          &     &  $g$-tensor  &     & & HFI Ga$_{(1)}$ & & &HFI Ga$_{(2)}$  &  & &HFI Mg/Zn \\
      &&&&&&$A$ (G)&&&$A$ (G)&&&$A$ (G)&  \\\hline

Mg$_\mathrm{Ga1}$ &          & 2.0096 & 2.0241 & 2.0205 &         &       &          & 14.98 & 15.26 & 14.76   & 3.26 & 3.21 & 2.54    \\
                  & $\theta$ &  90    &     0  &   90   &         &       &          & 85    & 41    & 49      & 90   & 0    & 90      \\
                  & $\phi$   &  14    &        &  -76   &         &       &          & -9    & 76    & 86      & -25  &      & 65      \\ \hline
Mg$_\mathrm{Ga2}$ &          & 2.0088 & 2.0222 & 2.0271 &   19.90 & 19.02 & 18.39    & 12.73 & 12.64 & 13.03   & 1.94 & 1.43 & 2.03    \\
                  & $\theta$ &   85   &   13   &  78    &   88    & 8     & 83       & 77    &  41   & 52      &  78  & 42   & 50      \\
                  & $\phi$   &  23    &  -45   &  -68   &   27    &       & -64      & -32   & 74    & 48      &  -8  & -84  & 72      \\ \hline 
Zn$_\mathrm{Ga1}$ &          & 2.0089 & 2.0078 & 2.0190 &         &       &          & 15.58 & 15.32 & 15.90   & 7.88 & 6.74 & 3.68    \\
                  & $\theta$ &  89    &     0  &   90   &         &       &          & 89    & 46    & 44      & 90   &  0   & 90      \\
                  & $\phi$   &  -1    &        &  -89   &         &       &          & -10   & 80    & 81      & -9   &      & 81      \\ \hline
Zn$_\mathrm{Ga2}$ &          & 2.0125 & 2.0207 & 2.0330 &   17.14 & 17.30 & 18.06    & 12.44 & 12.72 & 12.34   & 11.83 &  3.77 & 9.90  \\
                  & $\theta$ &  41    &    52  &   77  &     58  &  43   &  64      & 78    & 50    & 42       & 85   & 43    & 47  \\
                  & $\phi$   &   29   &     3   &  -77   &     -25 &  -73  &  48      & -21   & 78    & 55     & -8   & 87  & 77  \\ \hline
                  &          &  $a_*$ &   $b$  &   $c$  &   $a_*$ &   $b$ &   $c$    & $a_*$ & $b$   & $c$     &  &   &    \\ \hline
HSE\cite{Ho18}   &  &   &  &  &  27.2 & 28.9 & 25.7    & 14.4 & 14.7 & 14.2 \\\hline
Expt.\cite{KananenMg} & &          2.0038 & 2.0153 & 2.0371    &  26.1 & 25.6 & 25.5 & 11.8 & 11.9 & 11.3 
    \end{tabular}
  \end{ruledtabular}
\end{table*}

\section{Results}
\begin{figure*}
  \includegraphics[width=16cm]{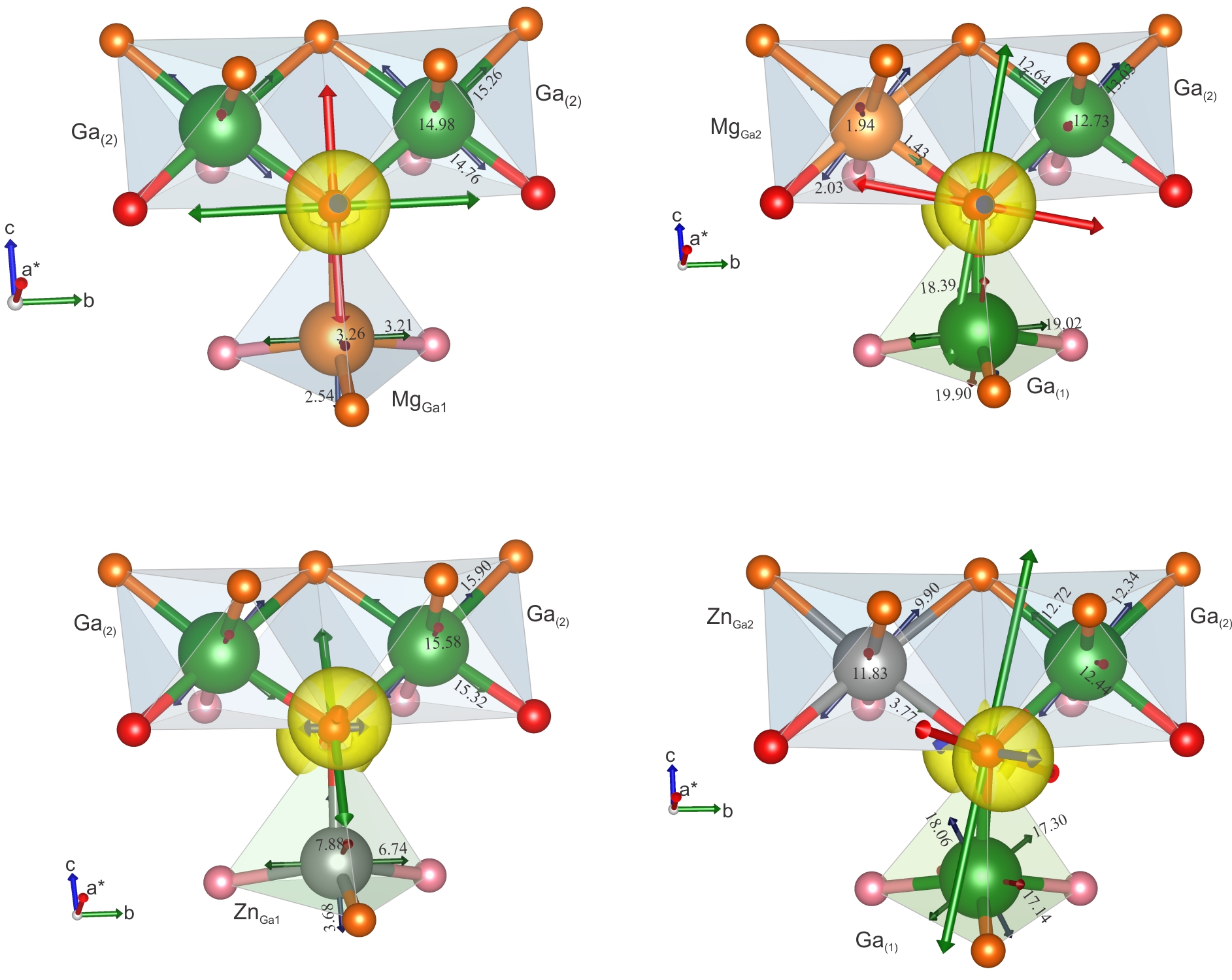}
  \caption{Mg$_\mathrm{Ga1}$, Mg$_\mathrm{Ga2}$, Zn$_\mathrm{Ga1}$, and Zn$_\mathrm{Ga2}$ structures, spin density in yellow, $g$-tensor principal 
    axes indicated by thick double arrows
    with length proportional to the $\Delta g$  (deviation from free electron value $g_e = 2.002391$), green colored Ga atoms
    are the ones with strong HFI. The small O spheres are color coded red O$_{(1)}$, pink O$_{(2)}$,
    orange O$_{(3)}$ and the polyhedra surrounding the Ga and their type are indicated. The thin double arrows show the principal axes of HFI.
  \label{Combfig}}
\end{figure*}

We consider Mg$_\mathrm{Ga1}$ (tetrahedral) as well as Mg$_\mathrm{Ga2}$
(octahedral) sites although previous work has already shown that
the octahedral site has lower energy and calculate their EPR relevant
properties in the neutral charge state. 
The calculated $g$-tensors principal values and axes as
well as the SHF splitting $A$ tensor are given in Table \ref{tabepr}.
Their corresponding spin-densities for both sites are
shown in Fig. \ref{Combfig} as yellow isosurfaces along with
the tensor principal axes, shown as arrows and the highlighted atoms
on which there is significant SHF interaction. 
In previous work \cite{KananenMg,Ho18} the $g$ tensor and $A$-tensor
values were presented along ${\bf a}$ instead of ${\bf a}_*$, but
strictly speaking the three principal axes should be orthogonal
to each other and the tensor is only fully specified by giving the
values along the three principal axes, and the directions of the latter. 
The calculated principal axes are close to but not
exactly along the mutually orthogonal
directions ${\bf a}_*$ (\ie the direction of the reciprocal lattice vector),
${\bf b}$ and ${\bf c}$.  The angle between ${\bf c}$ and ${\bf a}$ is
103.7$^\circ$, so ${\bf a}_*$ and ${\bf a}$ differ by 13.7$^\circ$ only, so
this does not make much  difference when comparing to the experimental values
the way they were specified.

We can see in Fig. \ref{Combfig} that for Mg$_\mathrm{Ga1}$ the
principal axes are very close to the crystal axes. However, for
Mg$_\mathrm{Ga2}$, which is slightly more asymmetric and tilted,
they are farther away from the crystal axes. 
The spin density is in both cases seen to be oriented close to ${\bf a}_*$
which is the direction in which the $\Delta g$ is smallest for both centers.


We can see that for the tetrahedral site, the spin density is well
localized on an O$_{(1)}$ next to the Mg on the mirror plane of the
crystal structure and with strong SHF interaction on two
equivalent octahedral Ga$_{(2)}$ neighbors with a nearly isotropic
SHF tensor $A$. The $A$-tensor principal axes  however are not oriented
close to the crystal axes but rather along the octahedral bonds, one of
which is along ${\bf a*}$ but the other are nearly 45$^\circ$ tilted from ${\bf b}$ and ${\bf c}$. 
In contrast, for the octahedral site, the spin density is still located on
O$_{(1)}$ but with hyperfine on two inequivalent Ga neighbors.
Again, the $A$-tensor in the Ga$_{(2)}$ neighbor have
their principal axes along the octahedral bonds. For The Ga$_{(1)}$ 
neighbor one axes is close to ${\bf c}$ along one of the tetrahedral bonds
while the other two axes also are close to the tetrahedral bond directions.

The octahedral Mg$_\mathrm{Ga2}$ site model
  is in agreement with experimental findings of
Kananen \etal\cite{KananenMg} while the tetrahedral Mg$_\mathrm{Ga1}$  is not. 
In terms of the $g$-tensor there is also good agreement with the
Mg$_\mathrm{Ga2}$ site. For Ga$_{(2)}$ the $g$-tensor
in the ``close to'' ${\bf c}$ direction
is calculated to be slightly higher than in the close to ${\bf b}$
direction, in agreement with experiment, while for  Mg$_\mathrm{Ga1}$
it is the other way around. However, these differences are almost
within the errorbars of calculating $\Delta g$ and it is really
the inequivalence of the SHF that is the telltale sign of
the octahedral site.

The assignment of the experimental EPR center to Mg$_\mathrm{Ga2}$
agrees with Ho \etal's \cite{Ho18} finding that the Mg has lower
total energy on the octahedral site. Nonetheless, it might not completely
exclude some Mg to also occur on tetrahedral sites if the
doping process has some non-equilibrium aspects and thus it might
be possible in the future to detect also the Mg$_\mathrm{Ga1}$ EPR center
whose properties are here predicted. In terms of the SHF agreement with
experiment, our values with GGA underestimate the experimental values
for the Ga$_{(1)}$ by 23\% and overestimate the ones for Ga$_{(2)}$
by 10 \% while Ho \etal\cite{Ho18} overestimates the   Ga$_{(1)}$
values by 6 \% and the  Ga$_{(2)}$ ones by 24 \%.  In terms of anisotropy
of $A$, neither calculation predicts the ordering of the tensor principal
axes components correctly as given by the experiment.
However, all calculations agree with experiment
that the anisotropy is small and hence more or less within the errorbar.
Furthermore, the  present calculations indicate that the hyperfine
tensors have principal axes along the bond direction rather than along
the crystal directions  but insufficient experimental detail on this
is presently given in the experimental paper.\cite{KananenMg}
Thus there is no clear advantage to the hybrid functional in predicting
the EPR parameters, provided the structure is properly relaxed with
spin well localized on a single O.


On both sites, this acceptor
was found to have a very deep $0/-$ transition level about equal
for both sites.\cite{Ho18} They are significantly less deep in GGA
than in hybrid functional. Kyrtsos \etal \cite{Kyrtsos18} find
the Mg acceptor level
in GGA at 0.26/0.22 eV above the VBM for tetrahedral and octahedral site respectively, while in the hybrid functional calculation, they lie at
1.25 /1.05 eV according to Kyrtsos \etal and at 1.62/1.57 eV according to
Ho \etal\cite{Ho18}.  The differences between these two hybrid functional
results may stem from different treatment of the image charge corrections,
where in one case\cite{Kyrtsos18} a static dielectric constant
was used for screening and in the other \cite{Ho18}  the high-frequency
value was used. Slighlty different hybrid functional parameters such as the
fraction ($\alpha$) of exact (non-local) exchange, screened with screening
length $\mu$, included
($\alpha=0.26$, $\mu=0$)\cite{Ho18,Deak17} vs. ($\alpha=0.32$, $\mu=0.2$ \AA$^{-1}$)\cite{Kyrtsos18} may also play a role.  
Thus the exact values of these transition levels are still under dispute
and have not yet been settled by experiment. 
 In any case these levels lie rather deep below the CBM in $\beta$-Ga$_2$O$_3$
  which is usually unintentionally n-type or at least semi-insulating, 
  and thus they can only be made
  EPR active by optical excitation removing the electron from the $q=-1$ state.

Next, we consider the Zn$_\mathrm{Ga1}$ and  Zn$_\mathrm{Ga2}$ acceptors.
According to Kyrtsos et al. \cite{Kyrtsos18} the $0/-$ transition  level
for this acceptor lies a bit higher above the VBM than for Mg, at 0.35/0.27 eV (GGA)
or 1.39/1.22 eV (hybrid) for the
tetrahedral/octahedral site. 
Again, we calculate their properties in the neutral state only, which
is the EPR active state.  As we hypothesized, we find that in this
charge state, the tetrahedral site has lower energy than the
octahedral site, in fact, by 0.16 eV.  
The resulting $g$-tensor and hyperfine tensors are given in Table \ref{tabepr}
and the spin densities and tensors are visualized in Fig.\ref{Combfig}.
These are very similar to the corresponding Mg case but, based on the total
energy, we predict in this case that the tetrahedral site should be
easier to find experimentally. It would be characterized by
hyperfine on two equivalent Ga atoms and a $g$-tensor with maximum value
along ${\bf c}$. For the Zn$_\mathrm{Ga2}$ site, the principal axes
of the tensor are again somewhat further from the crystal axes.
Interestingly, for Zn$_\mathrm{Ga1}$ the  $g$-tensor has its lowest
value along ${\bf b}$ even though the spin density is mostly along ${\bf a}_*$
but both values have in fact small deviations from the free electron value.
It is also worthwhile pointing out that the highest value of the $g$-tensor
in this case is along the ${\bf c}$ principal axis and two
relatively small values are found in the orthogonal directions. It is thus close
to the $g$-tensor previously assigned
to a $V_\mathrm{Ga_2}$, \cite{Jurgen18,Skachkov19} which also
has two nearly equivalent Ga atoms and with $A$ values close to the ones calculate here.
It might thus be difficult to distinguish the Zn$_\mathrm{Ga1}$ EPR center from the Ga$_{(2)}$
vacancy one. It is however highly unlikely that
the samples in the previous experimental studies of Ga-vacancies
induced by irradiation would have contained Zn and  the Ga-vacancy
EPR centers became only visible after high energy particle irradiation
which is required to create the vacancies in the first place, because
the vacancies have high energy of formation. 
It would be interesting to search for the here predicted Zn-related
EPR center in a sample doped with Zn and without irradiation.

Finally, we mention that  Mg and Zn in principle also could show hyperfine
splittings from the $^{67}$Zn (4.1\% abundance) and $^{25}$Mg (10 \% abundance) isotopes, both
corresponding to a $I=5/2$ nuclear spin. Because of the low abundances, these would
be difficult to detect but we nonetheless provide their hyperfine properties in
the last columns of Table \ref{tabepr}. One may see that for Zn, the
values are comparable to Ga and show a larger anisotropy.

\section{Conclusions}
In summary, we have calculated the EPR signatures of the Mg and
Zn acceptors in $\beta$-Ga$_2$O$_3$ for both candidate sites, the octahedral
and tetrahedral one. Based on total energy calculations, the tetrahedral
is predicted to be preferred for the Zn, while the octahedral site
is preferred for Mg.  Both sites are shown to have distinct
EPR signatures and the predictions agree well with experiment for
the Mg case. Our results also show that hybrid functional calculations
are not clearly providing improved  hyperfine splitting parameter results compared to 
GGA calculations.  Nonetheless some type of orbital dependent functiona, hybrid or
DFT$+U$ is required to obtain a correctly relaxed structure with spin localized on
one oxygen. In terms of the hyperfine tensors, we find that their principal axes
occur close to the octahedral and tetrahedral bond directions rather than along the
crystalline axes.

\acknowledgements{This work was supported by
  the National Science Foundation (NSF), Division of Materials  Research (DMR)
  under grant No. 1708593. The calculations were done at the Ohio Supercomputer Center under Project No. PDS0145.}

\bibliography{ga2o3,dft,gipaw,hyperfine,ldau}

\end{document}